\documentclass{rspublic}

\begin{document}

\def\MSUN{\rm M_{\odot}}
\def\RSUN{\rm R_{\odot}} 
\def\MSUNYR{\rm M_{\odot}\,yr^{-1}}
\def\MSUNS{\rm M_{\odot}\,s^{-1}}
\def\MDOT{\dot{M}}

\newbox\grsign \setbox\grsign=\hbox{$>$} \newdimen\grdimen \grdimen=\ht\grsign
\newbox\simlessbox \newbox\simgreatbox
\setbox\simgreatbox=\hbox{\raise.5ex\hbox{$>$}\llap
     {\lower.5ex\hbox{$\sim$}}}\ht1=\grdimen\dp1=0pt
\setbox\simlessbox=\hbox{\raise.5ex\hbox{$<$}\llap
     {\lower.5ex\hbox{$\sim$}}}\ht2=\grdimen\dp2=0pt
\def\simgreat{\mathrel{\copy\simgreatbox}}
\def\simless{\mathrel{\copy\simlessbox}}
\title[MHD collapsar model]{Magnetohydrodynamic simulations\\ 
of the collapsar model\\
for early and late evolution of gamma-ray bursts}

\author[D. Proga]{Daniel Proga}

\affiliation{Physics Department, University of Nevada, Las Vegas, 
Nevada 89154 , USA}

\label{firstpage}

\maketitle

\begin{abstract}{accretion, accretion discs -- MHD -- methods: 
numerical -- gamma-rays: bursts}
I present results from magnetohydrodynamic (MHD) 
simulations of a gaseous envelope collapsing onto a black hole. 
These results support the notion that the collapsar model is one of most 
promising scenarios to explain the huge release of energy in a matter 
of seconds associated with Gamma Ray Bursts (GRB). Additionally, 
the MHD simulations show that at late times, when the mass supply rate is 
expected to decrease, the region in the vicinity of the black hole can 
play an important role in determining the rate of accretion, its time 
behaviour, and ultimately the energy output. In particular, the magnetic flux 
accumulated around the black hole can repeatedly stop and then restart 
the energy release. As proposed by Proga and Zhang, the episode or 
episodes of reoccurring of accretion processes
can correspond to X-ray flares 
discovered recently in a number of GRBs.
\end{abstract}

\section{Introduction}
The collapsar model proposes that long duration gamma-ray bursts (GRBs) 
are powered by  accretion  onto the collapsed iron core of a massive star
(Woosley 1993 and Paczy\'{n}ski 1998). As gas accretes at a very high rate 
($\sim 1 \MSUNS$), large neutrino and magnetic fluxes, a powerful outflow, 
and a GRB are produced. This model is strongly supported by the association 
of long-duration GRBs with stellar collapse (e.g., Hjorth et al. 2003; 
Stanek et al. 2003) and by theoretical studies
of stellar collapse 
(MacFadyen \& Woosley 1999; Popham, Woosley \& Fryer 1999; Proga et al. 2003; 
Mizuno et al. 2004; Fujimoto et al. 2006).

Recent GRB observations obtained with {\it Swift} provide 
new challenges to the model as early X-ray afterglow lightcurves of
nearly half of the long-duration GRBs show X-ray flares
(Burrows et al. 2005; Romano et al. 2006; Falcone et al. 2006).
The model must then account for an initial powerful burst of energy
and for an extended activity in a form of X-ray flares.

\section{MHD simulations}

I present results from direction simulations
of early evolution of stellar collapse and explore 
the implications of these and other simulations
to infer the physical conditions in the vicinity of a black hole (BH)
during the late phase of evolution, i.e., when most
of the stellar mass is accreted.
To study the extended GRB activity, one would need to follow the collapse
of the entire star. However, such studies are beyond current computer
and model limits. 

\subsection{Early time evolution}

Proga et al. (2003) performed the first MHD simulation of a collapsar model.
Their simulation begins after the $1.7~\MSUN$ iron
core of a 25~$\MSUN$ presupernova star has collapsed and follows the
ensuing accretion of the $7~\MSUN$ helium envelope onto the central
BH formed by the collapsed iron core.  
A spherically symmetric progenitor model is assumed, but the symmetry is
broken by the introduction of a small, latitude-dependent angular
momentum and a weak split-monopole  magnetic field. 
The simulation includes pseudo-Newtonian potential, 
a realistic equation of state, photodisintegration of bound nuclei and cooling 
due to neutrino emission. 

The simulation shows that
the early phase of evolution starts with
a transient episode of infall. Then the rotating gas starts to pile up 
outside the BH and forms a thick accretion torus bounded by a centrifugal 
barrier near the rotation axis. 
Soon after the torus forms 
(i.e., within a couple of orbital times at the inner edge), 
the magnetic field is amplified by the magnetorotational
instability (MRI, e.g., Balbus \& Hawley 1998) and shear.
MRI facilitates outward transport of angular momentum in the torus 
and enables torus accretion onto a BH. 
Another important effect of magnetic fields is that the torus produces 
a magnetized corona and an outflow.  The presence of the corona and outflow 
is essential to the evolution of the inner flow at all times and 
to the entire flow close to the rotational axis during the latter phase 
of the evolution.
The outflow very quickly becomes sufficiently strong 
to overcome supersonically infalling gas and to escape the star.
The top left panel in Fig. 1 shows a schematic picture on the inner most 
part of the flow.
The main conclusion from the simulation is that,
within the collapsar model, MHD effects alone are able to launch,
accelerate and sustain a strong polar outflow.

Other important insights gained from the MHD simulation include:
1) In the accretion torus, the toroidal field dominates over the
poloidal field and the gradient of the former drives a polar outflow;
2) The polar outflow can be Poynting flux-dominated; 
3) The polarity of the toroidal field can change with time;
4) The polar outflow reaches
the outer boundary of the computational domain ($5\times10^8$~cm) with
an expansion velocity of 0.2~c; 
5) The polar outflow is in a form of a relatively narrow jet 
(when the jet breaks through the outer boundary its half opening angle is 
$5^\circ$); 
6) Most of the energy
released during the accretion is in neutrinos, $L_\nu=2\times
10^{52}~{\rm erg~s^{-1}}$. 
Neutrino driving will increase the outflow energy 
(e.g., Fryer \& M\'{e}sz\'{a}ros 2003
and references therein), but could also increase
the mass loading of the outflow if the energy is deposited in the torus.

Due to limited computing
time, Proga et al.'s (2003) simulation was stopped at $t=0.28215$~s, 
which corresponds to 6705 orbits of the flow near the inner boundary.
The total mass and angular
momentum accreted onto the BH during the simulation  are
$0.1~\MSUN$ and $3\times10^{39}~{\rm g~cm^2~s^{-1}}$, respectively.
Thus, the simulation captured just the beginning
of the accretion and one 
expects the accretion to continue much longer, roughly the collapse
timescale of the Helium core ($\sim 10$~s).
However, one also expects a decrease of the mass supply
rate with time, especially in the late phase of activity,
because the stellar mass density decreases with increasing radius.

The long time evolution of MHD accretion flows was studied by 
Proga \& Begelman (2003)
who explored simulations very similar to the one
in Proga et al. (2003) but with much simpler physics
(i.e., an adiabatic equation of state, no neutrino cooling
or photodisintegration of helium). Next, I present some key
results from Proga \& Begelman (2003) as they are relevant
to late evolution of a collapsing star.

\subsection{Late time evolution}

The early phase of the time evolution  and the dynamics of the innermost flow,
are very similar in simulations present in Proga et al. (2003) 
and Proga \& Begelman (2003). In particular, after an
initial transient behaviour, the flow settles into a complex convolution
of several distinct, time-dependent flow components
including an accretion torus, its corona and outflow, and
an inflow and outflow in the polar funnel (see 
the bottom left panel in Fig. 2).

However, the late evolution of Proga \& Begelman's 
(2003) simulations shows that the torus accretion can
be interrupted by a strong poloidal magnetic field in
the vicinity of a BH (compare the bottom left and right panels 
in Fig. 2).  Thus the region in the vicinity of the BH
and the BH itself can play an important role in determining
the rate and time behaviour of the accretion and the energy output.

Motivated by this result, Proga \&  Zhang (2006) conjectured that  
the energy release can be repeatedly stopped and then restarted, 
provided the mass supply rate decreases with time even if the decrease 
is smooth (see Fig. 1). 
Proga \& Zhang appealed to the fact that, as mass is being accreted onto
a BH, the magnetic flux is accumulating in the vicinity of the BH.
Eventually, this magnetic flux must become dynamically important
and affect the inner accretion flow, unless the magnetic field
is very rapidly diffused.
Analytic estimates derived by Proga \& Zhang
show that the  model can account for the observed features of the X-ray
flares.

\section{Conclusions}

Both numerical and theoretical models of magnetized accretion flows
support the notion that long duration GRBs can be explained
by the collapsar model. The models also show that the innermost
part of the flow and accretor can respond {\em actively} to
changes of the accretion flow at larger radii.
In particular, the innermost accretion flow can be halted
for a very long time as to account for  
an extended activity of GRBs with X-ray flares.

\begin{acknowledgements}
I acknowledge support from NASA under ATP grant NNG06GA80G
\end{acknowledgements}

\begin{figure}
\begin{picture}(180,350)
\put(-150,360){\includegraphics{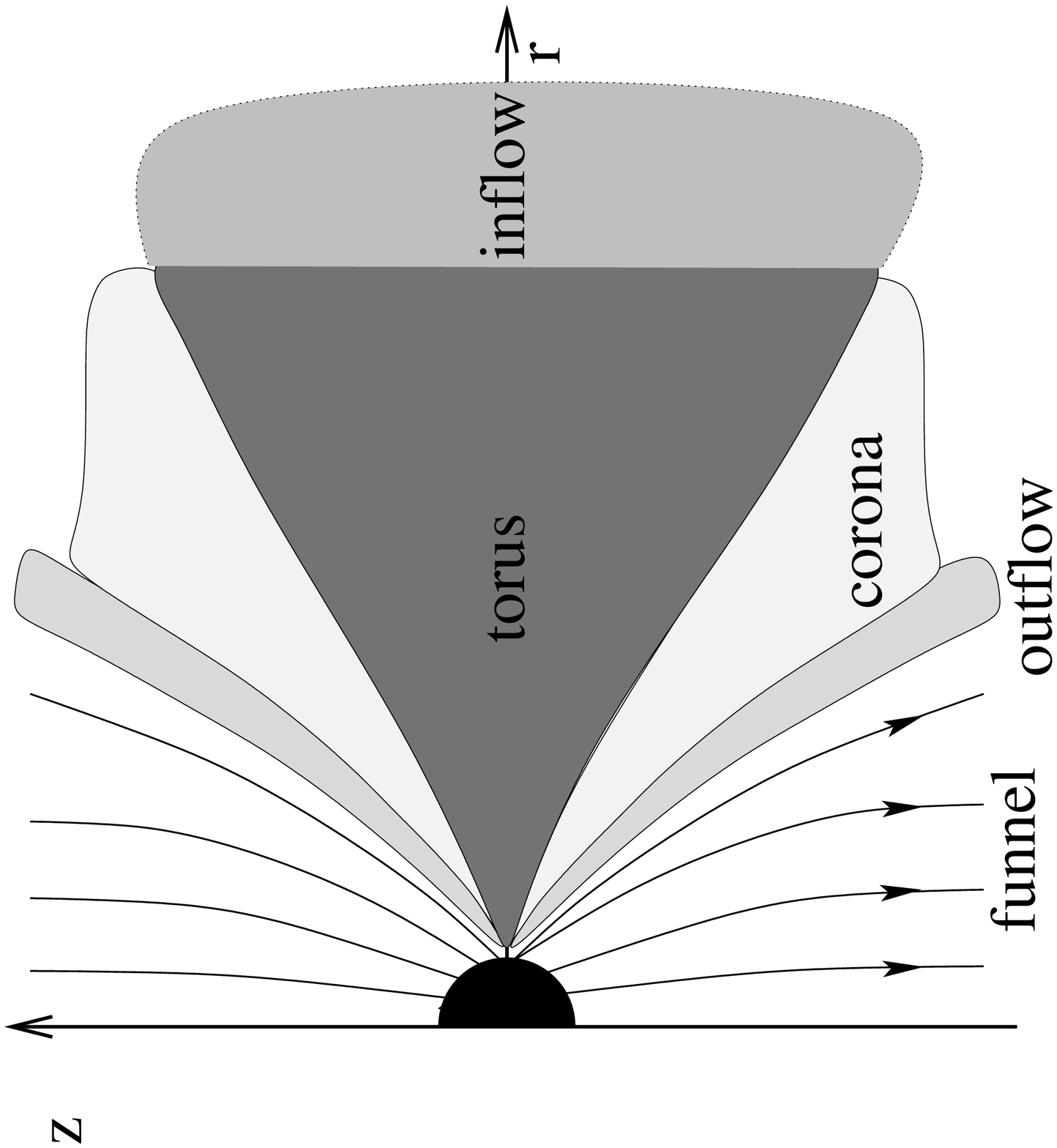}}
\put(50,360){\includegraphics{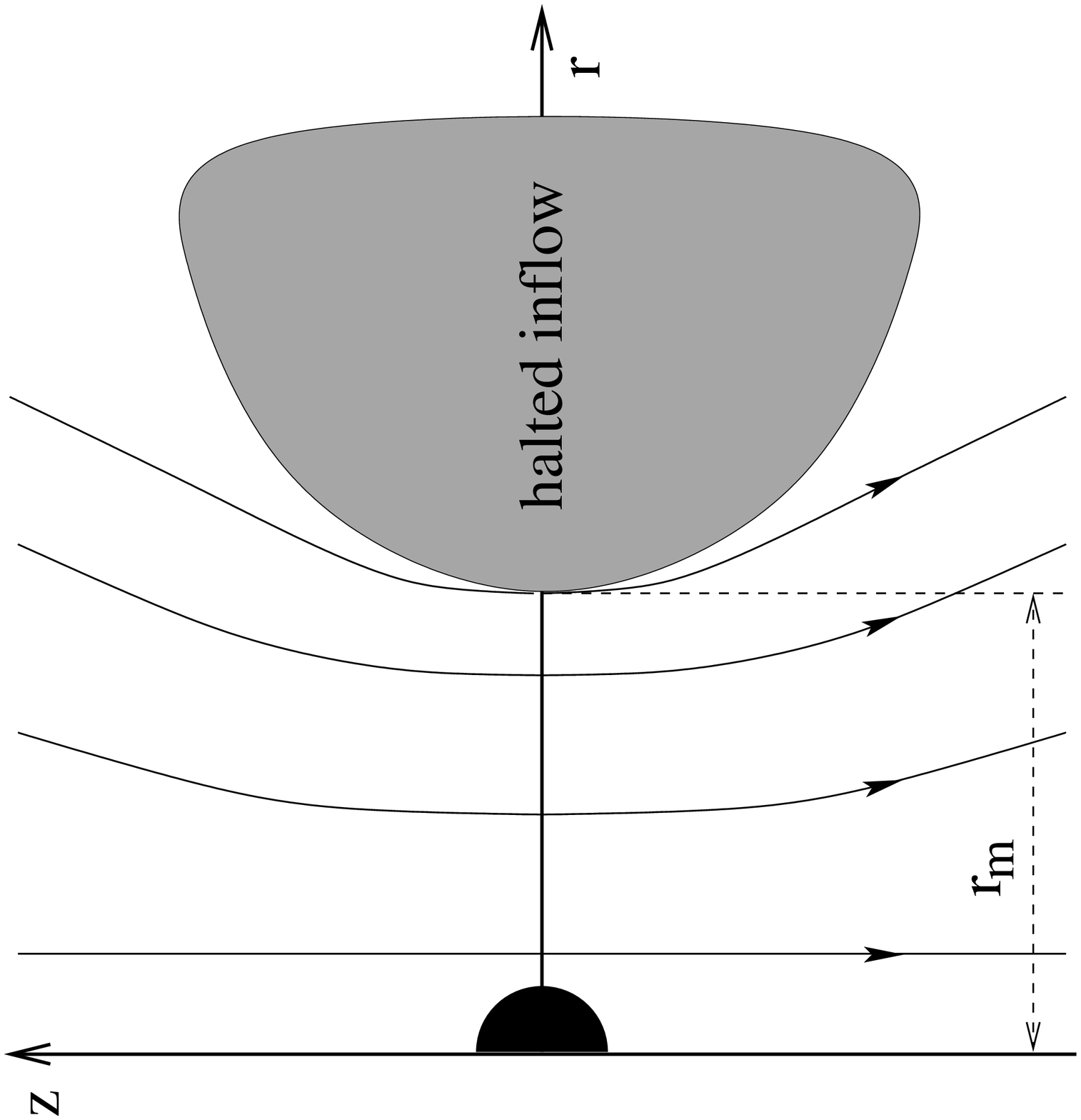}}
\put(-50,180){\includegraphics{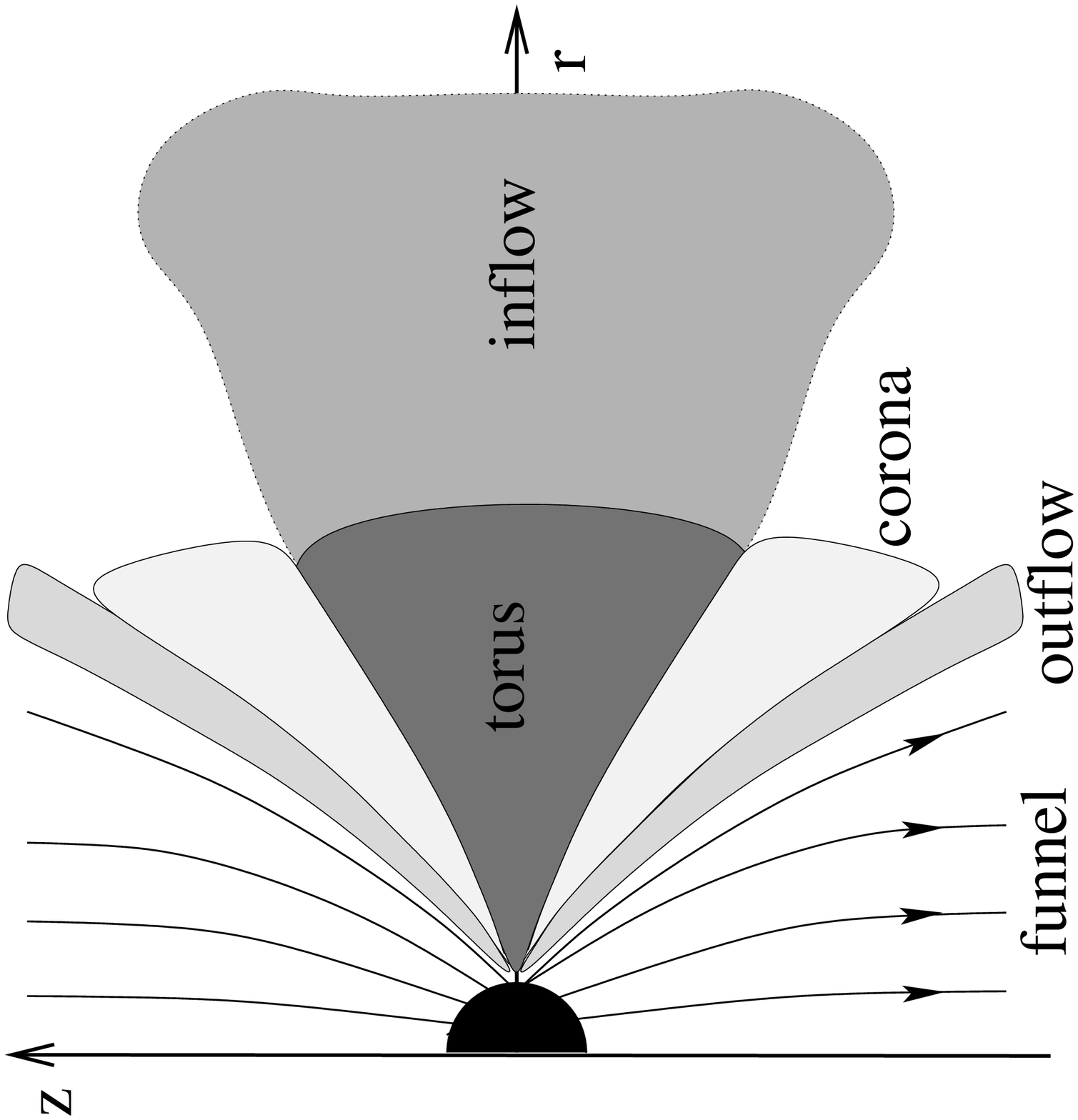}}
\end{picture}
\caption{\small
General structural features of the inner MHD flow
during three different accretion stages:
(1) The inner flow during the hyperaccretion (left top panel) when
a powerful jet forms and a strong poloidal magnetic
field is being accumulated at the center. The hyperaccretion 
can not be sustained because the mass supply rate from the outer inflow
drops with time (hence different shades used for the torus and inflow).
(2) The halted inflow (right top panel) when the hyperaccretion ended and 
the inflow rate is relatively low. During this stage the magnetic field 
accumulated earlier can support the gas against gravity. 
Consequently the inflow almost stops at the distance comparable to 
the magnetospheric radius, $r_m$. 
This stage ends when the surface density of the flow is too high 
for the magnetic field to support the gas.
(3) The inner flow when the magnetosphere is
squashed by the gas accumulated in the front of the inflow (bottom panel). 
The accretion torus is rebuilt and a powerful jet is regenerated.
The accretion rate at this stage is lower than
the hyperaccretion rate but higher than the inflow rate.
During the late accretion the inner flow
switches between the second and third stage and the third
stage corresponds to the time when X-ray flares are produced.
This cartoon illustrates the situation when the central magnetic flux
is conserved (i.e., the solid lines with arrows correspond to the magnetic
field lines at the center; for the clarity of the cartoon, the magnetic
field lines of other flow components are not drawn.)}
\end{figure}

\begin{figure}
\begin{picture}(180,480)
\put(-40,180){\includegraphics{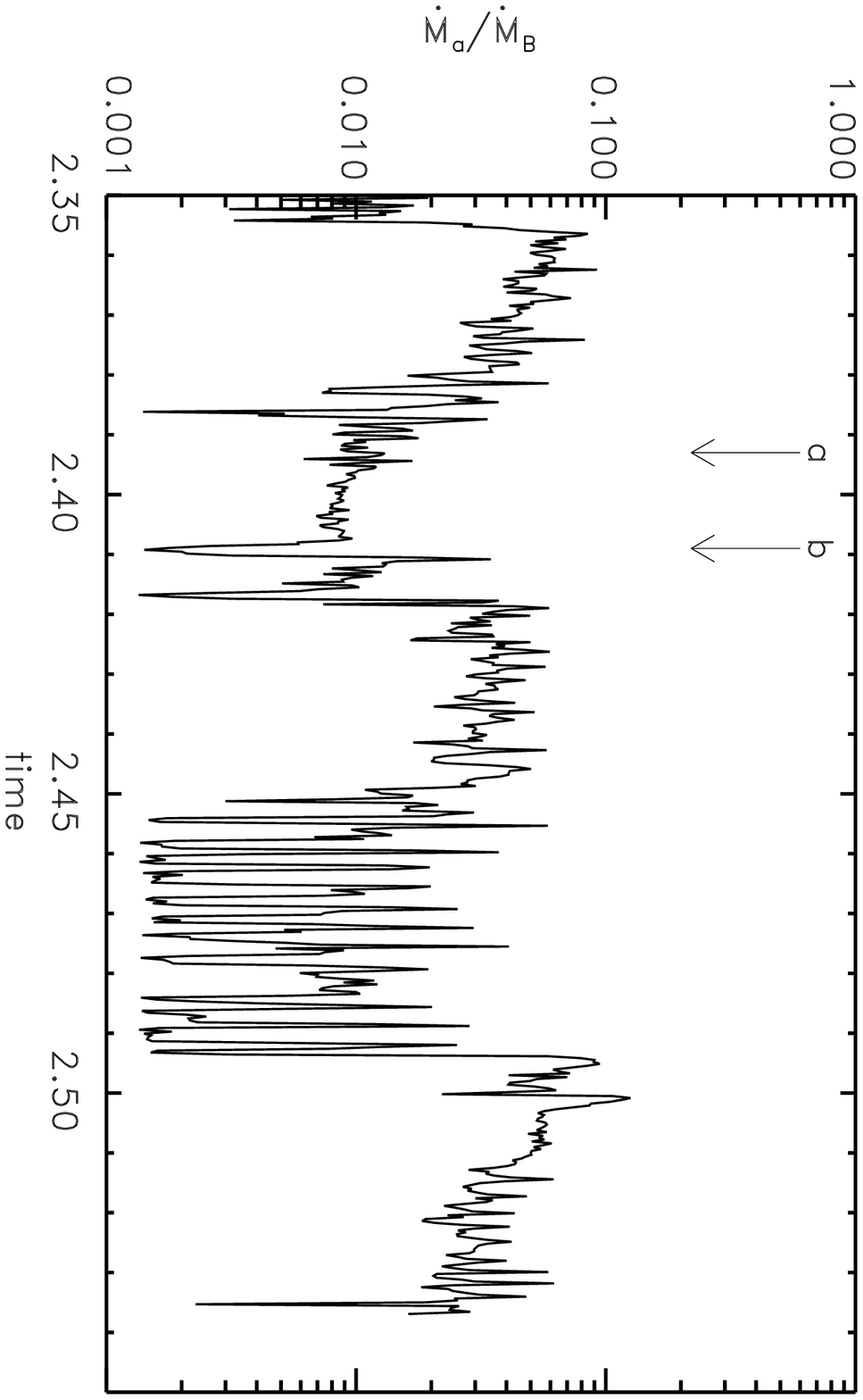}}

\put(180,30){\includegraphics{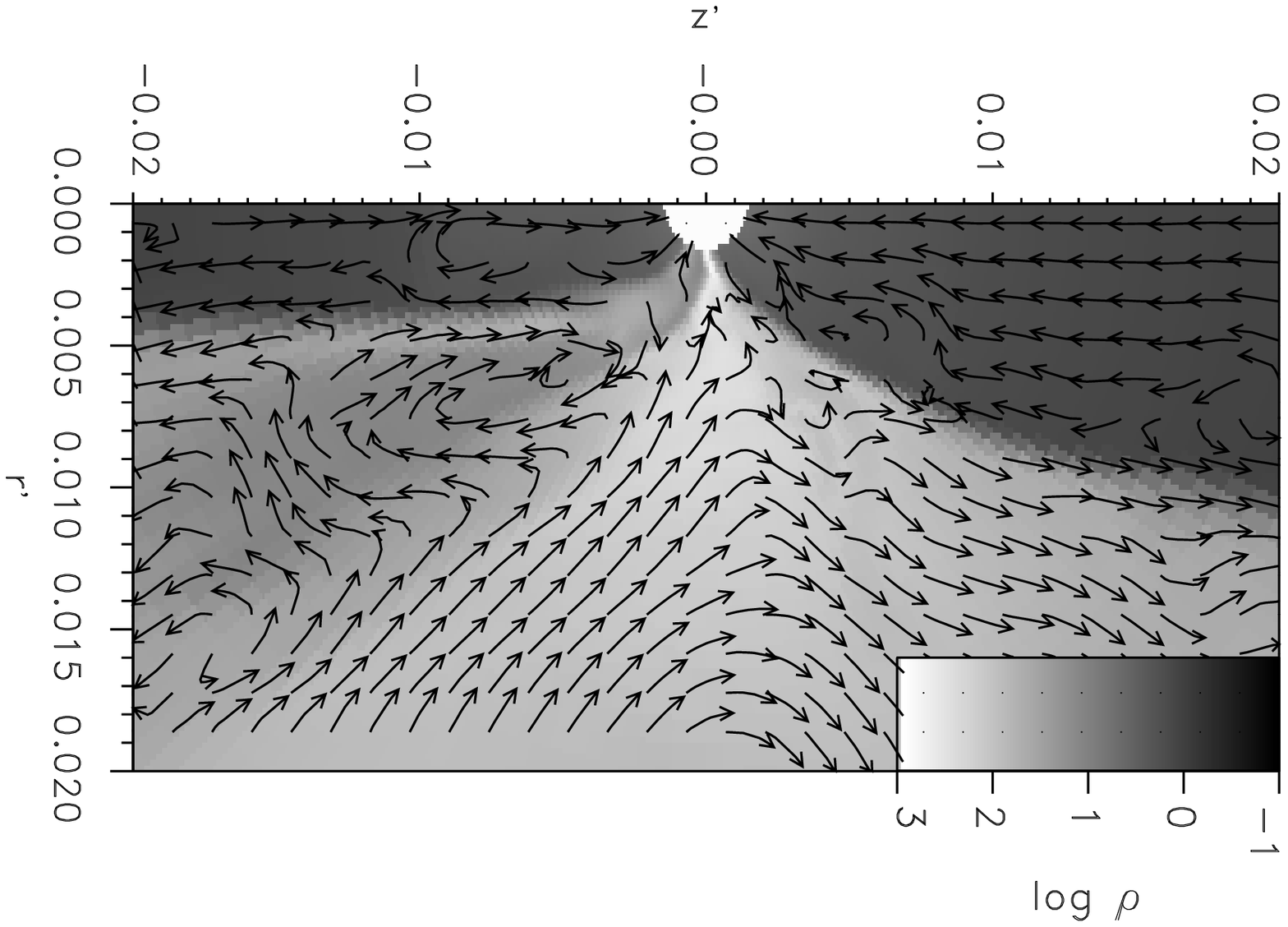}}
\put(380,30){\includegraphics{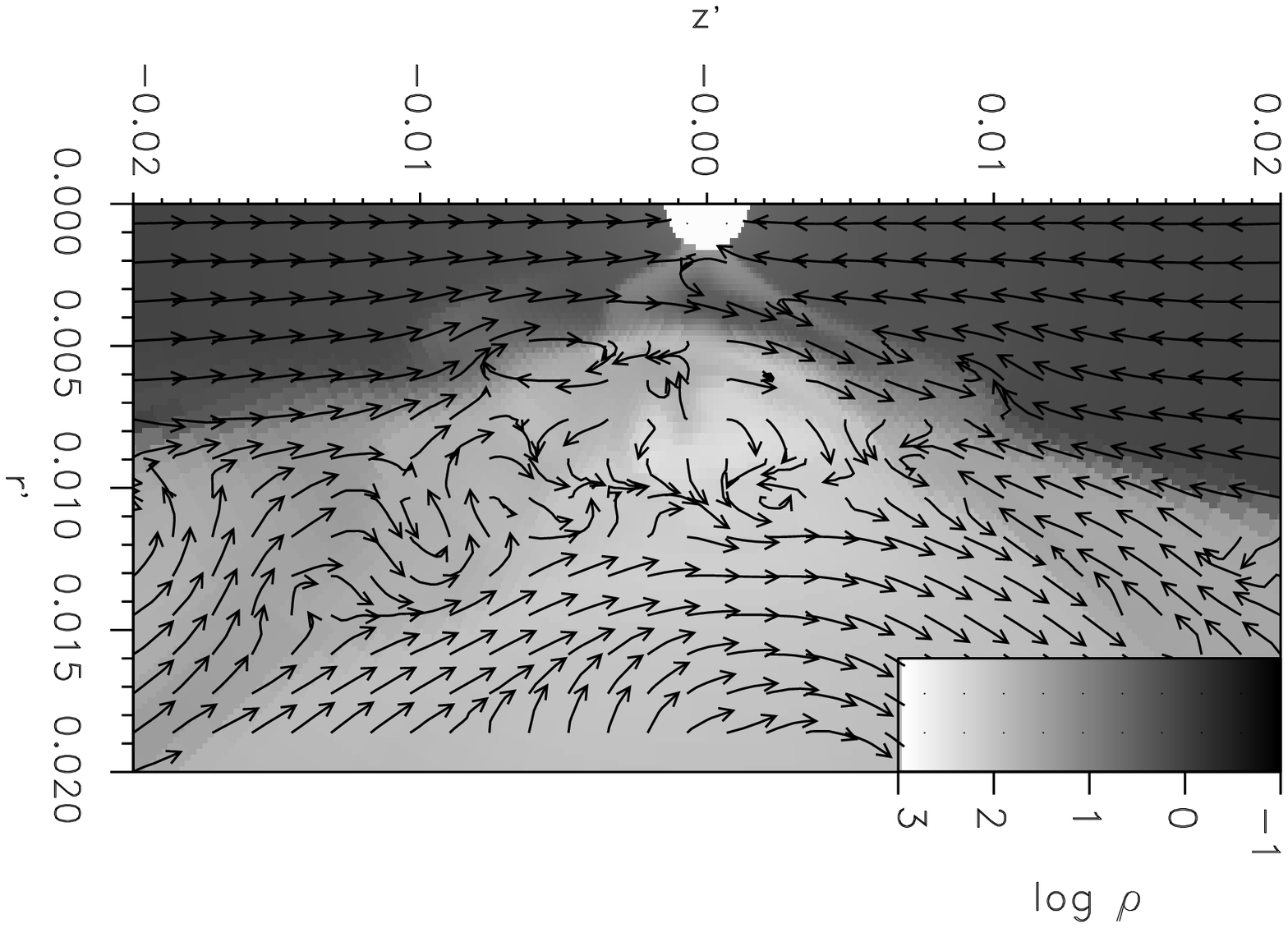}}
\end{picture}
\caption{{\it Top Panel:} 
Late time evolution of the mass accretion rate in units of the
Bondi rate, for Proga \& Begelman's (2003) run~D.
Vertical arrows mark times corresponding to the 
leff and right snapshots shown in the bottom row of panels 
(arrow a and b, respectively).
{\it Bottom Row of Panels:}
Maps of logarithmic density overplotted by the direction
of the poloidal velocity. These panels compare 
the inner flow in two different accretion states
mark by arrows in the top panel (see also fig. 1).
The time in the top panel is 
in units of the Keplerian orbital time at the Bondi radius, $R_B$.
The length scale and density in the bottom panels is
in the units of $R_B$ and the density at infinity, $\rho_\infty$ respectively.
In this simulation, the ratio between the Bondi radius and BH radius is 1200.
}
\end{figure}

\label{lastpage}

\end{document}